\documentclass[jgrga]{agu2001}
\usepackage{amsmath}
\usepackage{amssymb}
\usepackage{graphicx}
\usepackage{bm}

\journalid{}
\articleid{}{}
\paperid{}
\cpright{}{}
\received{} \revised{} \accepted{} \published{}

\authorrunninghead{SWISDAK ET AL.}
\titlerunninghead{TRANSITION GUIDE FIELD}
\authoraddr{M. Swisdak,
Icarus Research, Inc., PO Box 30780 
Bethesda MD 20824-0780, USA ,
(swisdak@ppd.nrl.navy.mil)}

\begin{document}

\title{The Transition from Anti-Parallel to Component Magnetic
Reconnection}

\authors{M. Swisdak, \altaffilmark{1}
J. F. Drake, \altaffilmark{2}
M. A. Shay\altaffilmark{2}
and J. G. McIlhargey, \altaffilmark{3} }

\altaffiltext{1}
{Icarus Research, Inc., Bethesda, MD, USA.}

\altaffiltext{2}
{IREAP, University of Maryland, College Park, MD, USA.}

\altaffiltext{3}{Department of Physics, University of
Maryland-Baltimore County, Baltimore, MD, USA.}

\begin{abstract}
We study the transition between anti-parallel and component
collisionless magnetic reconnection with 2D particle-in-cell
simulations.  The primary finding is that a guide field
$\approx\thinspace\negthickspace0.1$ times as strong as the asymptotic
reconnecting field --- roughly the field strength at which the
electron Larmor radius is comparable to the width of the electron
current layer --- is sufficient to magnetize the electrons in the
vicinity of the x-line, thus causing significant changes to the
structure of the electron dissipation region.  This implies that great
care should be exercised before concluding that magnetospheric
reconnection is antiparallel.  We also find that even for such weak
guide fields strong inward-flowing electron beams form in the vicinity
of the magnetic separatrices and Buneman-unstable distribution
functions arise at the x-line itself.  As in the calculations of {\it
Hesse et al.} [2002]\nocite{hesse02a} and {\it Yin and Winske}
[2003]\nocite{yin03a}, the non-gyrotropic elements of the electron
pressure tensor play the dominant role in decoupling the electrons
from the magnetic field at the x-line, regardless of the magnitude of
the guide field and the associated strong variations in the pressure
tensor's spatial structure.  Despite these changes, and consistent
with previous work, the reconnection rate does not vary appreciably
with the strength of the guide field as it changes between $0$ and a
value equal to the asymptotic reversed field.

\end{abstract}

\begin{article}

\section{\label{intro}Introduction}

The fast dissipation of magnetic energy in collisionless plasmas is a
common occurrence in nature, with examples ranging from tokamak
sawtooth crashes to magnetospheric substorms to solar flares.  The
process common to these phenomena is thought to be magnetic
reconnection, in which oppositely directed components of the magnetic
field cross-link, forming an x-line configuration.  The expansion of
the newly connected field lines away from the x-line converts magnetic
energy into kinetic energy and heat while pulling in new flux to
sustain the process.

Observations suggest that in many systems the ratio of the
characteristic reconnection time to the Alfv\'en crossing time is
$\sim 0.1$.  The simplest magnetohydrodynamic (MHD) description of
reconnection [{\it Sweet}, 1958; {\it Parker},
1957]\nocite{parker57b,sweet58a} is inconsistent with this value,
being too slow by several orders of magnitude [{\it Biskamp},
1986]\nocite{biskamp86a}.  However, the numerical simulations
comprising the GEM Reconnection Challenge [{\it Birn et al.},
2001]\nocite{birn01a} showed that the inclusion of the Hall effects,
which are important at small spatial scales and are neglected in MHD,
can produce fast reconnection.  The magnetic topology in these
simulations was understandably quite simple: equal and anti-parallel
fields separated by a thin current layer.  Yet even in the
magnetotail, where this approximation is often close to reality, a
small field directed parallel to the current (a guide field) is often
observed [{\it Israelovich et al.}, 2001]\nocite{israelovich01a}.

The effects of a guide field, $B_g$, on magnetic reconnection have
been examined before.  Sharp differences have been seen in the
large-scale flows around the x-line [{\it Hoshino and Nishida} 1983;
{\it Tanaka} 1995; {\it Pritchett} 2001]\nocite{hoshino83a,
tanaka95a,pritchett01b} as well as the pressure and magnetic field
signatures [{\it Kleva et al.}, 1995; {\it Rogers et al.}
2003]\nocite{kleva95a, rogers03a}.  Three-dimensional particle
simulations of similar systems without [{\it Zeiler et al.} 2002]
\nocite{zeiler02a} and with [{\it Drake et al.},
2003]\nocite{drake03a} a guide field showed that the former was
basically laminar in the direction parallel to the guide field while
the latter developed strong turbulence.  {\it Pritchett and Coroniti}
[2004] \nocite{pritchett04a} noted that moderate guide fields
($B_g/B_0 \lesssim 1$, where $B_0$ is the reconnecting field) have
only a slight effect on the reconnection rate, although {\it Ricci et
al.} [2004] \nocite{ricci04a} found somewhat slower rates for larger
fields ($B_g/B_0 = 3,5$). {\it Hesse et al.} [1999,2002]
\nocite{hesse99a,hesse02a} and {\it Yin and Winske}
[2003]\nocite{yin03a} showed that non-gyrotropic electron motions
balance the reconnection electric field at the x-line in both the
anti-parallel and guide field cases.

In light of these results, determining the minimum guide field $B_g$
that changes the structure of the x-line becomes of interest.  If it
satisfies $B_g > B_0$ then the effects of a guide field can usually be
ignored in the magnetosphere.  On the other hand, if the transition
occurs when $B_g << B_0$ guide-field reconnection is typical, and
anti-parallel is a special case perhaps only relevant in simulations.
We argue, based on both simulations and theoretical grounds, that the
transition occurs when the electron Larmor radius in the guide field
at the x-line becomes smaller than the width of the electron current
layer, ${\it i.e.}$ for $B_g/B_0 \approx 0.1$.  The implication is
that most magnetospheric reconnection is probably component
reconnection.

In section \ref{compute} of this paper we present our computational
scheme and initial conditions.  Section \ref{noguide} presents results
for the case $B_g=0$, section \ref{guide1} for $B_g = 1.0$ and section
\ref{transition} for $B_g = 0.2$.  We summarize our results and
discuss their implications for understanding magnetospheric
reconnection in section \ref{discussion}.

\section{\label{compute}Computational Details}

Our simulations are done with p3d, a massively parallel
particle-in-cell code [{\it Zeiler et al.}, 2002] \nocite{zeiler02a}.
The electromagnetic fields are defined on gridpoints and advanced in
time with an explicit trapezoidal-leapfrog method using second-order
spatial derivatives.  The Lorentz equation of motion for each particle
is evolved by a Boris algorithm where the velocity $\mathbf{v}$ is
accelerated by $\mathbf{E}$ for half a timestep, rotated by
$\mathbf{B}$, and accelerated by $\mathbf{E}$ for the final half
timestep.  To ensure that $\bm{\nabla \cdot} \mathbf{E} = 4\pi \rho $
a correction electric field is calculated by inverting Poisson's
equation with a multigrid algorithm.

The equations solved by the code are written in normalized
units. Masses are normalized to the ion mass $m_i$, the magnetic field
to the asymptotic value of the reversed field, and the density to the
approximate value at the center of the current sheet (see below).
Other normalizations derive from these: velocities to the Alfv\'en
speed $v_A$, lengths to the ion inertial length $c/\omega_{pi} = d_i$,
times to the inverse ion cyclotron frequency $\Omega_{ci}^{-1}$, and
temperatures to $m_i v_A^2$.

Our coordinate system is chosen so that the inflow and outflow for an
x-line are parallel to $\mathbf{\hat{y}}$ and $\mathbf{\hat{x}}$,
respectively.  The guide magnetic field and reconnection electric
field are parallel to $\mathbf{\hat{z}}$.  For comparison, our
$\mathbf{\hat{x}}$, $\mathbf{\hat{y}}$, and $\mathbf{\hat{z}}$ unit
vectors correspond to $-\mathbf{\hat{x}}$, $\mathbf{\hat{z}}$, and
$\mathbf{\hat{y}}$ in GSM coordinates.  The simulations presented here
are two-dimensional in the sense that out-of-plane derivatives are
assumed to vanish, {\it i.e.}, $\partial/\partial z = 0$.

The initial equilibrium comprises two Harris current sheets [{\it
Harris} 1962] \nocite{harris62a} superimposed on a ambient population
of uniform density.  The reconnection magnetic field is
$B_x=\tanh[(y-L_y/4)/w_0]- \tanh[(y-3L_y/4)/w_0]-1$, where $w_0 =
0.25$ and $L_y = 6.4$ are the half-width of the initial current sheets
and the box size in the $\mathbf{\hat{y}}$ direction respectively.
This configuration has two current sheets and allows us to use fully
periodic boundary conditions.  The electron and ion temperatures, $T_e
= 0.05$ and $T_i = 0.5$, are initially uniform as is the guide field
$B_g$.  Except for the background (lobe) population, which can have
arbitrary density (here $n_{\ell}=0.2$), pressure balance uniquely
determines the initial density profile. In this equilibrium the
density at the center of each sheet is $\approx 1.1$ at $t=0$. At
$t=0$ we perturb the magnetic field ($\tilde{B_x}/B_0 \approx 0.1$) to
seed x-lines at $(x,y) = (L_x/4,3L_y/4)$ and $(3L_x/4,L_y/4)$. 

To conserve computational resources, yet assure a sufficient
separation of spatial and temporal scales, we take the electron mass
to be $0.01$ and the speed of light to be $20$.  The domain measures
$6.4$ on a side and the grid has $1024\times 1024$ points, which
implies that there are $\approx\negmedspace 16$ gridpoints per
electron inertial length and $2$ per electron Debye length.  To check
for convergence we doubled the box size for one run (for $B_g = 0.2$)
and saw no significant variation in our results.  

The particle timestep is $6\times 10^{-4}$, or $0.12 \omega_{pe}$.
Our simulations follow $\sim 10^9$ particles and conserve energy to
better than $1$ part in $1000$.

\section{\label{noguide}Overview, $B_g = 0$}

Investigating the critical value of $B_g$ with multiple 3-D
simulations would entail a prohibitive computational expense, so we
instead performed a series of 2-D simulations that varied only in the
strength of the guide field.  The restricted dimensionality means that
many turbulent modes, including the Buneman instability seen by {\it
Drake et al.} [2003]\nocite{drake03a}, are not present. However, other
investigators have found that the gross morphological features of
reconnection x-lines are roughly invariant in the direction parallel
to the current density [{\it Pritchett and Coroniti},
2004]\nocite{pritchett04a}.

\begin{figure}
\noindent\includegraphics{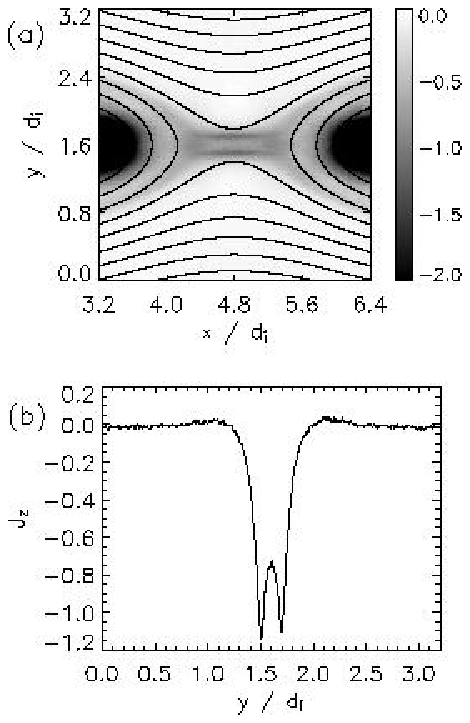}
\caption{\label{jznoguide} Reconnection in a system with $B_g = 0$ at
$t = 4.5$.  (a) The out-of-plane current density overlaid with
magnetic field lines in a region surrounding the x-line. The blackest
regions are not color-coded correctly, having been over-exposed to show
details near the x-line (b) A vertical cut through the x-line at
$x=4.8$. }
\end{figure}

A snapshot of the out-of-plane current density near the x-line for a
simulation with $B_g=0$ is shown in Figure \ref{jznoguide}.  The
initial current sheet has completely reconnected; the plasma at the
x-line at the time shown was in the low density ($n_{\ell}=0.2$) lobe
at the simulation's beginning.  Since $|\mathbf{B}|=0$ at the x-line
for $B_g = 0$ inward-flowing electrons must at some point find
themselves in a region where the magnetic field is too weak to
dominate their motion.  This happens at a distance from the x-line
given roughly by the electron inertial length, $c/\omega_{pe} \equiv
d_e$, which for this simulation is $\sqrt{m_e/n_{\ell}} \approx 0.2$
(in normalized units).  Once the electrons demagnetize they stream
towards the x-line parallel to the $\mathbf{\hat{y}}$ axis, through
the region of low magnetic field, until they turn due to the
increasing field on the opposite side of the current layer and reverse
direction.  They then execute ``figure-8'' trajectories [{\it Speiser}
1965]\nocite{speiser65a}, oscillating in the $\mathbf{\hat{y}}$
direction until escaping from the ends of the layer.  At the turning
points of their trajectories (where $v_y \rightarrow 0$), the local
electron density increases, forming a bifurcated current sheet.
Zeiler et al. [2002] have previously reported this bifurcation,
although it is particularly noticeable in our simulations because of
the high spatial resolution (16 gridpoints per $d_e$) and large ion to
electron temperature ratio ($T_i/T_e = 10$).  The bifurcation would be
obscured in simulations where these parameters had smaller values.
Note that this bifurcation is at a much smaller scale than the
ion-scale splits reported in Cluster observations [{\it Runov et al.} 
2003].\nocite{runov03a}

The bifurcation is also evident in Figure \ref{tnoguide}, which shows
the diagonal components of the electron temperature.  In analogy with
the definition of the fluid pressure tensor we define the electron
temperature tensor in a grid cell as a sum over the $N$ local
particles:
\begin{equation}\label{ttensor}
T_{\alpha \beta} = \frac{m_e}{N}\sum_{i=1}^N
(v_{\alpha,i}-<\negmedspace v_{\alpha}
\negmedspace>)(v_{\beta,i}-<\negmedspace v_{\beta}
\negmedspace >) \mbox{,}
\end{equation}
where $<\negmedspace\ldots\negmedspace>$ denotes an average,
$\it{e.g.,} <\negmedspace x \negmedspace > = \frac{1}{N}\sum_{i=1}^N
x_i$. Like the pressure tensor, which is related to the temperature
tensor by $P_{\alpha \beta} = nT_{\alpha \beta}$ where $n$ is the
density, the temperature is symmetric, $T_{\alpha \beta} = T_{\beta
\alpha}$.  For an isotropic plasma the off-diagonal elements of the
temperature tensor vanish while the diagonal elements are equal to
each other and to the scalar temperature, $T_{\alpha \alpha} = T_e$.
This is the case at $t=0$ in our simulations.

The decrease in $T_{yy}$ and $T_{zz}$ during inflow are consistent
with the adiabatic invariance of the magnetic moment $\mu$: $B \approx
B_x$ decreases, while $\mu \propto v_\perp^2/B$ remains constant.
$T_{xx}$, approximately the parallel temperature, simultaneously
increases due to energy conservation.  Any energy change due to the
interaction of the reconnection electric field $E_z$ and the curvature
and grad-$B$ drifts is small everywhere except near the x-line.

Once inside the layer the electrons demagnetize and, as previously
seen by {\it Zeiler et al.} [2002]\nocite{zeiler02a}, the electron
distribution in $v_y$ space separates into two counter-propagating
beams due to the cross-current layer bounce motion (see Figure
\ref{tnoguide}d).  As a consequence $T_{yy}$ sharply increases, as has
previously been noted by {\it Horiuchi and Sato}
[1997]\nocite{horiuchi97a}.

Despite the beams we see no evidence of a two-stream instability,
probably because of the small current layer width.  Unstable
wavenumbers for the electron-two-stream instability satisfy $k_yv_0 <
2\omega_{pe}$ where $v_0$ is the separation of the beam velocities
[{\it Krall and Trivelpiece} 1986]\nocite{krall86a}.  The maximum
growth rate, $\gamma = \omega_{pe}/\sqrt{8}$, occurs for $k_yv_0 =
\sqrt{3/2}\:\omega_{pe}$, and as $v_0 \rightarrow 0$ the growth rate
vanishes.  In the simulation the beam separation is largest at the
x-line, $v_0 \approx 8$, and drops to $0$ at the edges of the layer
$\approx 0.22 (= d_e)$ upstream.  With a local $\omega_{pe} \approx
90$ the instability criterion implies that only wavelengths $\lambda
\gtrsim 0.28$ are unstable at the x-line, and thus that the two-stream
instability is not excited in the narrow current layer.

\begin{figure}
\noindent\includegraphics{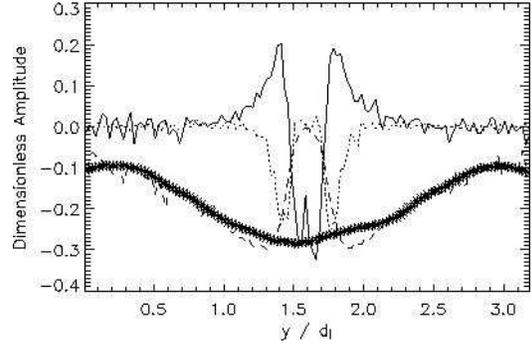}
\caption{\label{eznoguide} For $B_g=0$ a cut through the x-line at x
= $4.8$ showing the various terms balancing $E_z$ in equation
(\ref{ezbalance}).  $E_z$ is shown by the stars and the dashed, solid,
and dotted lines denote the $\mathbf{v \times B}$, divergence of the
pressure tensor, and inertial terms, respectively.  To reduce the
noise the plotted quantities were averaged over 20 gridpoints in x and
4 in y.}
\end{figure}

In a 2-D collisionless plasma the reconnection electric field at the
x-line is ultimately balanced by the divergence of the electron
pressure tensor [{\it Vasyliunas} 1975]\nocite{vasyliunas75a}.  In our
units the collisionless electron fluid momentum equation is
\begin{equation}
\mathbf{E} = -\mathbf{v_e} \mathbf{\times B} -
\frac{1}{n_e}\mathbf{\nabla \cdot} \mathbf{\Bar{\Bar{P}}_e} -
m_e (\mathbf{v_e \cdot} \mathbf{\nabla}) \mathbf{v_e} -
m_e\frac{\partial \mathbf{v_e}}{\partial t}
\end{equation}
which is exact insofar as the pressure tensor
$\mathbf{\Bar{\Bar{P}}_e}$ incorporates all kinetic effects not
included in the other terms. 
The reconnection electric field is thus
\begin{equation}\label{ezbalance}
\begin{split}
E_z &= -(v_x B_y-v_yB_x) \\
&- \frac{1}{n}\left(\frac{\partial
P_{xz}}{\partial x} + \frac{\partial P_{yz}}{\partial y}\right) \\
&-m\left(v_x\frac{\partial{v_z}}{\partial x} +
v_y\frac{\partial{v_z}}{\partial y} + \frac{\partial
v_z}{\partial t}\right)
\end{split}
\end{equation}
where we have dropped the electron subscript and used the fact that
$\partial/\partial z = 0$.  In a steady state only the pressure terms
can balance $E_z$ at the x-line, as can be seen in Figure
\ref{eznoguide}.  Far from the current layer the EMHD relation
$\mathbf{E} = -\mathbf{v_e \times B}$ holds, while nearer the x-line
both the off-diagonal elements of the pressure tensor and the inertial
terms are important.  At the x-line the pressure tensor terms
dominate.  Both $\partial P_{yz}/\partial y$ and $\partial
P_{xz}/\partial x$ contribute, although the former is larger in our
simulation by a factor of $\approx 2$.  The term proportional to
$\partial/\partial t$ is not shown separately but, as expected during
quasi-steady reconnection, is negligible.

\section{\label{guide1}Guide Field, $B_g = 1.0$}

A large guide field changes the structure of the x-line by both
lowering the total plasma $\beta$ and magnetizing the electrons
throughout the current sheet.  When $B_g$ is small the dominant wave
mode at small lengthscales, and hence the governor of the particle
dynamics, is the whistler.  This is seen when electrons in the outflow
region are accelerated by $E_z$ and drag the magnetic field out of the
reconnection plane [{\it Mandt et al.}, 1994; {\it Shay et al.},
1998]\nocite{mandt94a,shay98a}, causing the well-known quadrupolar
symmetry in $B_z$ along the separatrices [{\it Sonnerup}, 1979; {\it
Terasawa}, 1983] \nocite{sonnerup79a,terasawa83a}. As $B_g$ increases
the importance of the kinetic Alfv\'en mode grows [{\it Rogers et
al.}, 2001]\nocite{rogers01a}.  For that mode the coupling occurs when
$E_{||}$ accelerates electrons along newly reconnected field lines,
increasing the electron density on one side of the current layer,
decreasing it on the other, and forming a quadrupolar pattern [{\it
Kleva et al.}, 1995]\nocite{kleva95a}.  The perturbations in $B_z$
acquire a component determined by pressure balance, leading to a
symmetric component that can, for very large $B_g$, overwhelm the
quadrupolar pattern [{\it Rogers et al.}, 2003]\nocite{rogers03a}.
Figure \ref{bzne} shows the electron density and $B_z$ from the
simulation discussed in the previous section ($B_g = 0$) and one that
is otherwise identical except that $B_g=1.0$.

\begin{figure}
\noindent\includegraphics{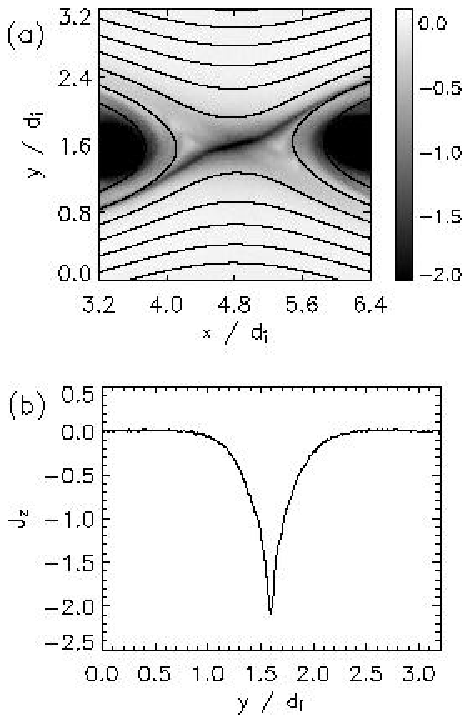}
\caption{\label{jzguide1} Reconnection in a system with $B_g = 1.0$ at
$t = 4.5$.  (a) The out-of-plane current density overlaid with
magnetic field lines in a region surrounding the x-line. Black areas
have been over-exposed to show detail at the x-line. (b) A vertical cut
through the x-line at $x=4.8$.}
\end{figure}

The parallel velocity of the electrons, mostly directed out of the
reconnection plane, develops a quadrupolar symmetry opposite to that
of the density (high density paired with low velocity and {\it vice
versa}).  The density asymmetry is a larger effect, however, and the
result is an out-of-plane current density that is canted with respect
to the initial current sheet, as can be seen in Fig.~\ref{jzguide1}.
Because inflowing electrons remain magnetized in the guide field they
do not have figure-8 trajectories at the x-line and the bifurcations
in the electron density and current density disappear.

\begin{figure}
\noindent\includegraphics{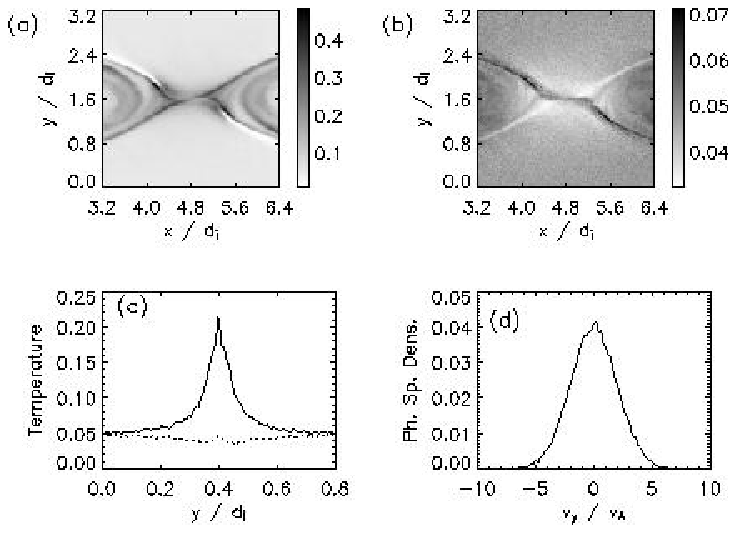}
\caption{\label{tguide1} Data for $B_g=1.0$. Panels (a) and (b) show
$T_{||}$ and $T_{\perp}$, respectively.  Panel (c) shows vertical cuts
at $x=4.8$.  The solid line is $T_{||}$ and the dashed $T_{\perp}$.
Panel (d) shows the distribution of $v_{ey}$ in a region measuring
$1.0\medspace d_e$ long and $0.5\medspace d_e$ high and centered on
the x-line.}
\end{figure}

In the $B_g=0$ simulation discussed in Section \ref{noguide} the
magnetic field on the inflow axis ($x = 4.8$) was dominantly parallel
to $\mathbf{\hat{x}}$ (except at the x-line where $|\mathbf{B}| = 0$).
For $B_g = 1.0$, in contrast, the field rotates $\approx 45^{\circ}$
between the lobe plasma and the x-line.  The rotation complicates the
interpretation of the temperature tensor of equation (\ref{ttensor}),
so to simplify we transform to a coordinate system where the axes are
parallel and perpendicular to the local magnetic field.  In this frame
\begin{equation}\label{parperp}
\mathbf{\Bar{\Bar{T}}} = [T_{||} \mathbf{\hat{b}\hat{b}} +
T_{\perp}(\mathbf{\Bar{\Bar{I}}}-\mathbf{\hat{b}\hat{b}})] +
\mathbf{\Bar{\Bar{T}}_{ng}}
\end{equation}
where $\mathbf{\hat{b}}$ is a unit vector in the direction of the
magnetic field, $\mathbf{\Bar{\Bar{I}}}$ is the unit tensor, and
$\mathbf{\Bar{\Bar{T}}_{ng}}$ contains the non-gyrotropic terms.
Images of the parallel and perpendicular temperatures along with cuts
through the x-line are shown for $B_g=1.0$ in Figure \ref{tguide1}a-c.

Conservation of magnetic moment again explains the decrease in
$T_{\perp}$ along the inflow direction.  Since $B_g$ is constant the
decrease in $B$ is smaller than was the case in section \ref{noguide},
and the relative decrease of $T_{\perp}$ in Figure \ref{tguide1}c is
smaller than in Figure \ref{tnoguide}d.  Because the electrons remain
magnetized at the x-line, $T_{\perp}$ remains small, in sharp contrast
to the results shown in Figure \ref{tnoguide}.  The increase in
$T_{||}$ inside the current sheet is due to the intermixing of colder
inflowing electrons and electrons accelerated by the parallel electric
field along the separatrices.  The unimodal distribution function of
Figure \ref{tguide1}d confirms that the electrons do not execute
Speiser-like orbits.

\begin{figure}
\noindent\includegraphics{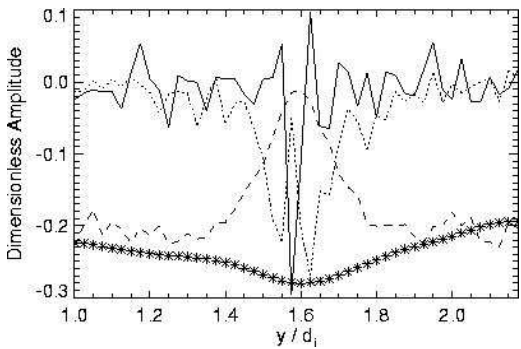}
\caption{\label{ezguide1} The terms balancing the reconnection
electric field for $B_g=1.0$.  $E_z$ is shown by the stars and the
dashed, solid, and dotted lines denote the $\mathbf{v \times B}$,
divergence of the pressure tensor, and inertial terms, respectively.
Note the difference in horizontal scale between this figure and Figure
\ref{eznoguide}.  To reduce noise the plotted quantities were averaged
over 16 gridpoints in x and 4 in y.}
\end{figure}

The terms balancing the reconnection field in equation
(\ref{ezbalance}) are shown for this case in Figure \ref{ezguide1}.
At the x-line the off-diagonal elements of the pressure tensor again
make the primary contribution although, as a comparison of Figures
\ref{ezguide1} and \ref{eznoguide} makes clear, the scale length over
which they are important is much smaller than when $B_g=0$.  This is
consistent with the results of {\it Hesse et al.} 
[2002]\nocite{hesse02a}.  Unlike the anti-parallel case the $\partial
P_{xz}/\partial x$ term makes the dominant contribution while the
$\partial P_{yz}/\partial y$ term is negligible.

\section{\label{transition}Transition}

For what $B_g$ does the transition between the reconnection of Section
\ref{noguide} and that of Section \ref{guide1} occur?  Far from the
x-line, where both species are completely magnetized, the guide field
cannot play an important dynamical role unless $B_g \gtrsim B_0$.  As
a particle approaches the current layer, however, the reconnecting
component decreases and the influence of the guide field rises.
Qualitatively, $B_g$ is important when the associated electron Larmor
radius is equal to the spatial scale associated with the x-line.

Consider a system with $B_g=0$ and examine the z component of the
electron equation of motion under quasi-steady conditions,
\begin{equation}\label{deltay}
m_e v_{ey}\frac{\partial v_{ez}}{\partial y} = -eE_z -
\frac{e}{c}(v_{ex}B_y - v_{ey}B_x) -\frac{1}{n}\frac{\partial
P_{yz}}{\partial y}\mbox{,}
\end{equation}
where we have assumed that derivatives with respect to $x$ can be
neglected when compared to those with respect to $y$.  In a 2D
steady-state system Faraday's Law implies that $E_z$ is relatively
uniform (see Figures \ref{eznoguide} and \ref{ezguide1}).  Far from
the x-line electrons are frozen to the magnetic field and the $E_z$
and $\mathbf{v \times B}$ terms are roughly equal.  Within the current
layer the convective part of the inertial term and the pressure tensor
become important, with the transition occurring at some lengthscale
$\Delta_y$ where the terms balance.  If, for simplicity, we restrict
our attention to the vertical axis through the x-line, symmetry
implies that $B_y$ is zero and
\begin{equation}\label{vzdywx}
v_{ez} = \Delta_y \Omega_{x,\text{up}}
\end{equation}
where $\Omega_{x,\text{up}} = eB_{x,\text{up}}/m_ec$ is the cyclotron
frequency based on the reconnecting field at $y = \pm \Delta_y$.  In
general $B_{x,\text{up}}$ will be less than the asymptotic
reconnecting field.  Within this inner scale the electrons carry most
of the current, so we also have
\begin{equation}\label{jz}
\frac{4 \pi}{c} J_z = \frac{4 \pi}{c} n v_{ez} \approx \nabla \times
\mathbf{B} \approx \frac{\partial B_x}{\partial y}
\end{equation}
where we have ignored both the displacement current and the
contribution from the current due to $\partial B_y/\partial x$.
Converting derivatives with respect to $y$ to division by $\Delta_y$
and combining equations (\ref{vzdywx}) and (\ref{jz}) we find that
\begin{equation}\label{delty}
\Delta_y = \frac{c}{\omega_{pe}} = d_e
\end{equation}
The electron velocity producing the current is $v_{ez} \sim d_e
\Omega_{x,\text{up}}$.

Now consider the addition of a small ambient guide field.  The
magnetic field no longer vanishes at the x-line and the electron
Larmor radius there is
\begin{equation}\label{rhol}
\rho_g = \frac{v_{ey}}{\Omega_{e,g}}
\end{equation}
where we have taken $v_{ex} \approx 0$ based on symmetry
considerations, $\Omega_{e,g} = eB_g/m_ec$ is the electron cyclotron
frequency based on $B_g$, and $v_{ey}$ is the electron inflow velocity
into the unmagnetized region around the x-line.  The guide field will
be important when this Larmor radius is smaller than the width of the
current layer, $\rho_g < \Delta_y$. 

The major contributors to the electron inflow velocity $v_{ey}$ are
the thermal speed and the $\mathbf{E} \times \mathbf{B}$ drift.  A
Sweet-Parker-like scaling suggests that the ratio of the $\mathbf{E}
\times \mathbf{B}$ inflow speed to the outflow speed is equal to the
normalized reconnection electric field $E_z$.  For a wide range of
conditions it has been shown [{\it Shay et al.}, 1999; {\it Shay et
al.}, 2001]\nocite{shay99a,shay01a} that the outflow is roughly equal
to the electron Alfv\'en speed $v_{Ae}$ and $E_z \approx 0.1$.  In the
low-temperature limit ($v_{th} << v_{E\times B}$) this argument
implies that $v_{ey}\approx 0.1v_{Ae}$ and the bound for a dynamically
important $B_g$ is given by
\begin{equation}
\rho_g = \frac{v_{ey}}{\Omega_{e,g}} = \frac{0.1 v_{Ae}}{\Omega_{e,g}}
< \Delta_y = d_e
\end{equation}
or
\begin{equation}\label{criteria}
B_g > E_z \sim 0.1.
\end{equation}
In our simulations $E_z$, and hence the transitional value of $B_g$,
is $\approx 0.2$.  

Equation (\ref{criteria}) is not valid when the electron thermal speed
$v_{the}$ dominates the contribution from the $\mathbf{E} \times
\mathbf{B}$ drift.  In the high-temperature limit one must
substitute $v_{th}$ rather than $v_{E\times B}$ for $v_{ey}$ in
equation (\ref{rhol}), and the relevant criterion becomes $\beta_g <
1$ where $\beta$ is evaluated with the lobe density and temperature.
In the simulations presented here $v_{the} \approx v_{E\times B}$ and
so the critical value of the guide field remains $B_g \approx 0.2$.

We note that the estimate for the electron inflow velocity
$v_{ye}\approx 0.1c_{Ae}$ is smaller than the counter-streaming
velocity of the electrons shown in Fig.~\ref{tnoguide}(d). This is
because a local electrostatic field $E_y$ develops inside the electron
current layer that accelerates the electrons towards the magnetic
null. However, this field decelerates the electrons once they cross
the null so this electric field does not change the effective electron
Larmor radius of the electrons in the guide field.

We have explored the transition from anti-parallel to finite guide
field reconnection through a series of simulations with $B_g = 0$,
$0.1$, $0.2$, $0.4$, and $0.75$ and found, in agreement with our above
arguments, that simulations with $B_g = 0.2$ most clearly display
characteristics intermediate between $B_g = 0$ and $B_g = 1$.

\begin{figure}
\noindent\includegraphics{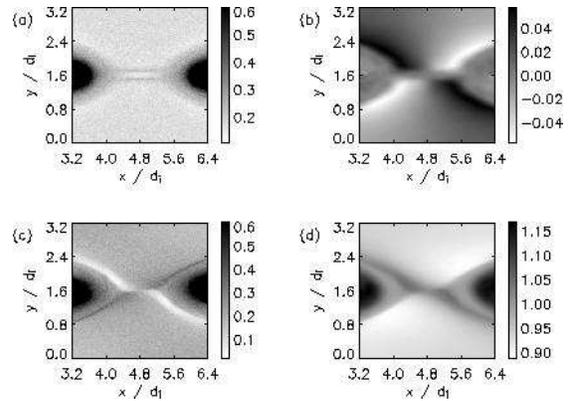}
\caption{\label{bzne} The electron density and out-of-plane magnetic
field $B_z$ in simulations with $B_g = 0$ and $1.0$. (a) $n_e$, $B_g =
0$; (b) $B_z$, $B_g = 0$; (c) $n_e$, $B_g = 1.0$; (d) $B_z$, $B_g =
1.0$.  In all panels $t = 4.5$.  In panels (a) and (c) black areas
have been over-exposed to show detail at the x-line. }
\end{figure}

\begin{figure}
\noindent\includegraphics{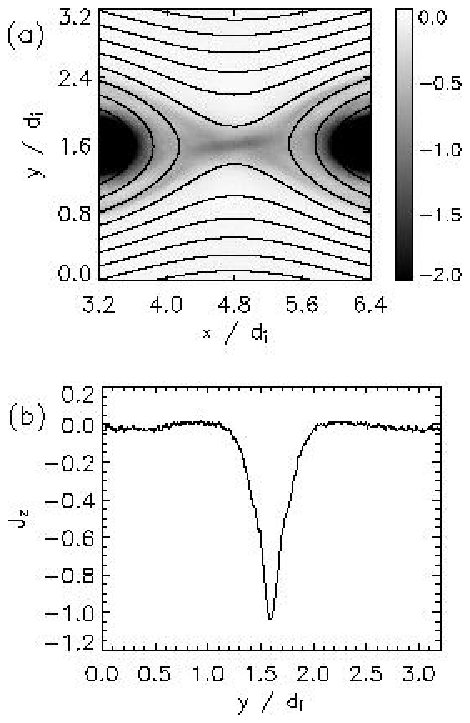}
\caption{\label{jzguide02} Reconnection in a system with $B_g = 0.2$
at $t = 4.5$.  (a) The out-of-plane current density overlaid with
magnetic field lines in a region surrounding the x-line. Black areas
have been over-exposed to show detail at the x-line. (b) A vertical cut
through the x-line at $x=4.8$.}
\end{figure}

\begin{figure}
\noindent\includegraphics{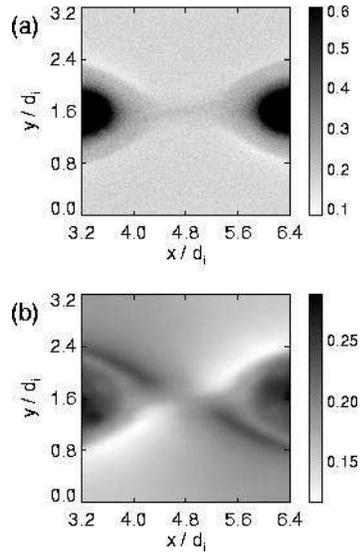}
\caption{\label{bzne02} The electron density (a) and out-of-plane magnetic
field $B_z$ (b) in a simulation with $B_g = 0.2$.  In panel (a) black
areas have been over-exposed to show detail at the x-line. }
\end{figure}

\begin{figure}
\noindent\includegraphics{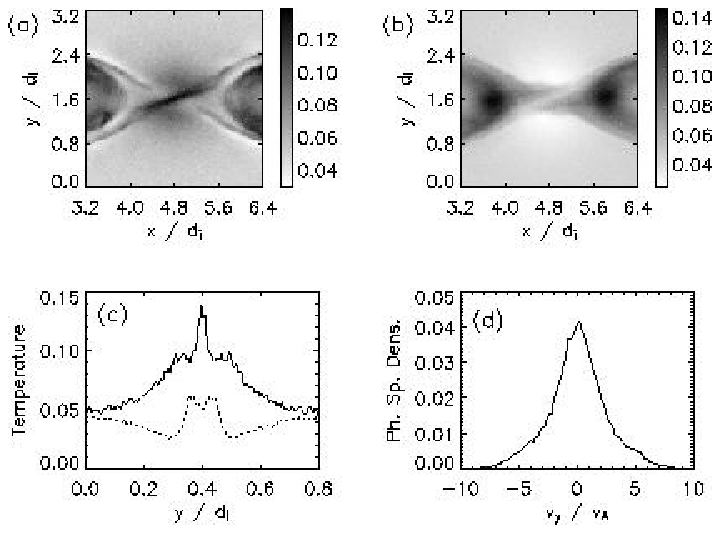}
\caption{\label{tguide02} Data from $B_g=0.2$. Panels (a) and (b) show
$T_{||}$ and $T_{\perp}$, respectively.  Panel (c) shows vertical cuts
at $x=4.8$.  The solid line is $T_{||}$ and the dashed $T_{\perp}$.
Panel (d) shows the distribution of $v_{ey}$ in a region measuring
$1.0\medspace d_e$ long and $0.5\medspace d_e$ high and centered on
the x-line.}
\end{figure}

Figure \ref{jzguide02} shows the out-of-plane current density for a
run identical to those previously discussed except that $B_g = 0.2$.
There is no bifurcation in the current density and the canting of the
current layer, while present, is not as strong as the case with $B_g =
1.0$ (Figure \ref{jzguide1}).  Figure \ref{bzne02} shows the electron
density and $B_z$ for this simulation.  The electron density is not
bifurcated and bears some resemblance to the quadrupolar cavities so
prominent in Figure \ref{bzne}c, while $B_z$, although still basically
quadrupolar, no longer has the strong symmetry obvious of Figure
\ref{bzne}b.  Evidence for a transition can also be seen in the
parallel and perpendicular temperatures shown in Figure
\ref{tguide02}.  The results are clearly intermediate between Figures
\ref{tnoguide} and \ref{tguide1}.  The cut in Figure \ref{tguide02}c 
shows that the increase in the parallel temperature as electrons
approach the x-line is similar in both magnitude and profile to the
$B_g = 0$ case.  Within the current layer, however, $T_{||}$ increases
to a sharp peak similar to that for $B_g=1$.  The perpendicular
temperature decreases towards the x-line and then, within the layer,
rises to a peak.  This peak is midway in magnitude between the $B_g=0$
and $B_g=1$ cases.

\begin{figure}
\noindent\includegraphics{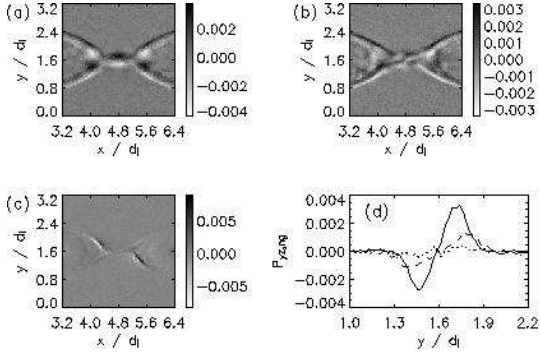}
\caption{\label{offp} Panels (a)--(c) show the non-gyrotropic
component of $P_{yz}$ for $B_g = 0, 0.2,$ and $1.0$. Panel (d) shows a
cut at $x=4.8$ in each figure.  The solid, dashed, and dotted lines
are for, respectively, $B_g=0, 0.2,$ and $1.0$.}
\end{figure}

In order to examine the variation with guide field of the off-diagonal
pressure tensor terms of Equation \ref{ezbalance} it is necessary to
separate the gyrotropic and non-gyrotropic contributions.  The
gyrotropic part is strongly influenced by the presence of a guide
field and, in any case, does not contribute to balancing $E_z$ at the
x-line.  Figure \ref{offp} shows the change in the non-gyrotropic
portion of $P_{yz}$ as the guide field varies.  The cuts in Figure
\ref{offp}d demonstrate that as $B_g$ increases the role of $\partial
P_{yz}/\partial y$ in balancing the reconnection electric field at the
x-line decreases dramatically.

\begin{figure}
\noindent\includegraphics{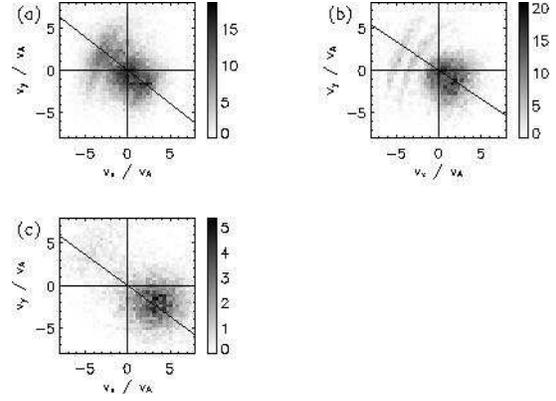}
\caption{\label{dist} Panels (a)--(c) show the $v_x$-$v_y$
distribution functions taken just upstream of the separatrix for runs
with $B_g = 0,0.2,1.0$, respectively.  The slanted line in each panel
shows the direction of the local magnetic field.  The colors are
measured in units of phase space density.  The counter-propagating
features at the upper left of each figure are due to electrons
accelerated at the x-line}
\end{figure}

The presence of a small guide field also has signatures far from the
x-line.  Figure \ref{dist} shows the $v_x$-$v_y$ distribution
functions for three simulations with different guide fields taken at
the same point, just upstream of the upper-left separatrix and, for
the runs with a finite guide field, inside the density cavity.
Because of the low plasma density within the cavity the parallel
electric field remains finite over an extended region along the
separatrix [{\it Pritchett and Coroniti}, 2004]\nocite{pritchett04a})
in the $B_g = 1.0$ system.  This electric field locally accelerates
the electrons, producing a strong beam flowing towards the x-line.
However even for $B_g = 0.2$ the beam is already clearly present.
(The features in the upper left quadrant of each panel are electrons
that have already been accelerated at the x-line).  There is a net
current but no distinct beam for $B_g = 0$.  The origin, detailed
structure, and effect of this extended region of $E_{||}$ is discussed
in a future publication.

\section{\label{discussion}Discussion}

This study suggests that only a minimal guide field, $B_g \approx 0.1
B_0$ is required to alter the dynamics of electrons both in the
vicinity of the x-line and at remote locations along the separatrices.
The implication is that in most real systems, including the
magnetotail, the guide field might not be negligible.  In any case
this study suggests that one can not simply ignore the guide field if
$B_g\ll B_0$.

\begin{figure}
\noindent\includegraphics{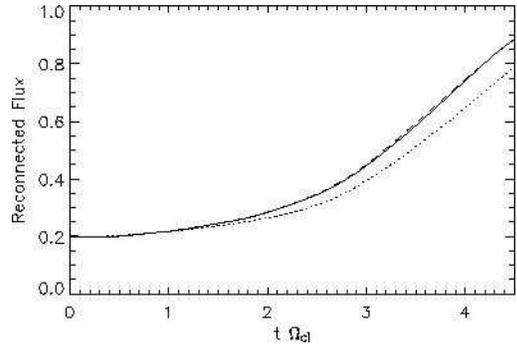}
\caption{\label{ratecomp} The reconnected flux versus time for the
three runs discussed in this work.  The solid, dashed, and dotted
lines correspond to $B_g = 0$, $0.2$, and $1.0$, respectively.  The
reconnection rate is the slope of the curves.}
\end{figure}

A counter-argument could be made that reconnection with $B_g \neq 0$
is significantly slower compared to the case with $B_g=0$ and
therefore the magnetosphere will self-select locations where the guide
field is nearly zero (less than $0.1$ of the anti-parallel field). We
find that this is not the case. The guide field alters the dynamics
both locally and at large scales significantly before the rate of
reconnection is significantly affected. Specifically, Figure
\ref{ratecomp} shows that magnetic flux reconnects only slightly ($\approx
10$\%) slower for $B_g = 1.0$ than for $B_g = 0$, a result consistent
with other simulations [{\it Rogers et al.}, 2003; {\it Pritchett and
Coroniti}, 2004]\nocite{pritchett04a,rogers01a}.  It should be noted,
however, the the onset of reconnection in real systems could be biased
either for or against guide fields.  Our simulations do not address
this question since they start with a finite perturbation that
effectively places the system in the nonlinear regime at $t=0$.

Another factor potentially affecting guide field reconnection is the
effect of an ambient pressure gradient and the associated diamagnetic
drifts.  At the magnetopause, where density gradients perpendicular to
the current layer produce diamagnetic drifts, the reconnection rate
can be strongly reduced [{\it Swisdak et al.},
2003]\nocite{swisdak03a}.  However, it was shown that diamagnetic
suppression occurs for small guide fields with the transition
occurring when $\beta_x > B_g/B_0$ (for density length scales of order
an ion inertial length).  Combined with the results of this work the
conclusion is again that magnetopause reconnection always includes a
dynamically important guide field.

\begin{figure}
\noindent\includegraphics{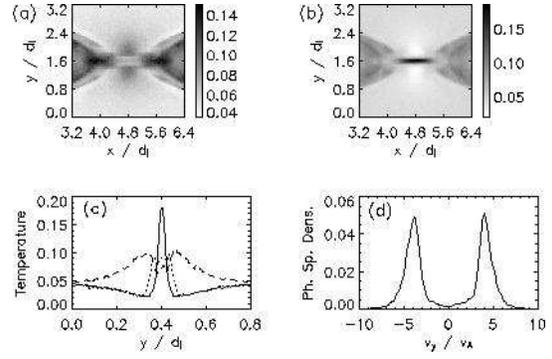}
\caption{\label{tnoguide} Data for $B_g=0$. Panels (a) and (b) show
the electron temperatures $T_{xx}$ and $T_{yy}$ (see the text for
definitions) near the x-line of Figure \ref{jznoguide}.  Panel (c)
shows cuts of the temperatures at $x =4.8$.  The solid line is
$T_{yy}$, the dashed $T_{xx}$, and the dotted $T_{zz}$.  Panel (d)
shows the distribution of $v_{ey}$ in a region measuring $1.0
\medspace d_e$ long and $0.5\medspace d_e$ high and centered on the
x-line.}
\end{figure}

Guide fields also play an important role in the development of
turbulence in three-dimensional reconnection simulations.  Simulations
by {\it Drake et al.} [2003]\nocite{drake03a} with $B_g = 5.0$ showed
that turbulence can self-consistently develop at a reconnection
x-line.  The acceleration of electrons by the reconnection electric
field led to a separation of the ion and electron drift speeds, which
then triggered the Buneman instability.  At late time the nonlinear
evolution led to the formation of electron holes, localized bipolar
regions of electric field.  These structures produced an effective
drag between the ions and electrons that was large enough to compete
with the off-diagonal pressure tensor in balancing the reconnection
electric field.  Earlier 3D simulations with $B_g = 0$ by {\it Zeiler
at al.} [2002]\nocite{zeiler02a} produced no significant turbulence at
the x-line once reconnection was established.  The strong electron
heating for $B_g = 0$ (as shown in Figure \ref{tnoguide}) suppressed
all streaming instabilities near the x-line since for $T_i \gg T_e$
such instabilities require a beam velocity $v_b$ greater than the
electron thermal speed $v_{te}$.  Yet since $B_g=5.0$ is significantly
larger than typical magnetospheric values it was unclear from these
studies whether turbulence and enhanced ion-electron drag were common
features of reconnection in the magnetosphere.

Our 2D simulations cannot produce the Buneman instability at the
x-line seen in these earlier simulations. However, we can examine the
distribution functions produced by our simulations and determine
whether they would be unstable in a full 3D system.  In the low
temperature limit with $\mathbf{k}$ parallel to both $\mathbf{B}$ and
the relative drift velocity $\mathbf{v_b}$ a plasma is Buneman
unstable to wavenumbers satisfying the relation $kv_b <\omega_{pe}$
[{\it Krall and Trivelpiece}, 1986]\nocite{krall86a}.  For finite
temperature plasmas with Maxwellian distributions the condition for
instability is more complicated and, in fact, for small drifts in warm
plasmas no instability exists.  The instability threshold for
arbitrary distributions can be found numerically, but a rough rule of
thumb is that a plasma is Buneman unstable if the electron and ion
velocity distribution functions do not substantially overlap ($v_b
\gtrsim v_{te}$).

\begin{figure}
\noindent\includegraphics{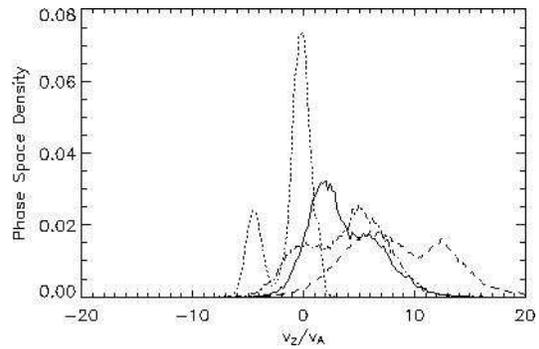}
\caption{\label{xdists} The $v_z$ distribution functions for three
runs.  The solid line shows the case $B_g = 0$, the dash-dotted $B_g =
0.2$ and the dashed $B_g=1.0$.  The dotted line shows the ions.}
\end{figure}

Figure \ref{xdists} shows the distribution of $v_z$ at the x-line for
runs with three values of the guide field, $B_g=0,0.2,1$.  The dotted
line shows the ion distribution and the double peak is a remnant of
the two initial populations, the non-drifting background and the
drifting population comprising the initial current sheet.  The three
electron distribution functions demonstrate that as the guide field
increases the x-line electrons acquire a larger drift with respect to
the ions.  The electron energy gain is limited by the amount of time
they spend within the current layer before advecting into the outflow
region.  Larger simulations suggest that our small system size may
prevent the electrons from reaching their maximum speeds.  A finite
$B_g$ acts like a guide wire, restraining this flow and leaving more
time for a particle to be accelerated by the reconnection electric
field.  If our simulation were three-dimensional the $B_g=1$ case
would almost certainly be Buneman unstable, while the $B_g=0$ case
would not.  Again, the $B_g=0.2$ case is transitional.

The small size of the transition guide field has important
implications for magnetospheric reconnection.  The x-line current
sheet should usually be canted with respect to the ambient current
layer.  Density and flow velocities measured on the separatrices
should have a quadrupolar symmetry.  Distribution functions taken just
upstream of the separatrices should exhibit an inward-flowing beam.
Turbulence, and electron holes in particular, should be common.
Magnetospheric reconnection with negligible guide fields should be
rare.

\begin{acknowledgments}
This work was supported in part by the NASA Sun Earth Connection
Theory and Supporting Research and Technology programs and by the NSF.
\end{acknowledgments}

\pagebreak

\end{article}

\end{document}